\documentclass[aps,prl,
reprint,
notitlepage,
superscriptaddress
]{revtex4-1}

\usepackage{amsmath,amssymb,amsfonts,mathrsfs}
\usepackage{bm}
\usepackage{graphicx}
\usepackage{hyperref}

\begin{document}
\title{Spectral Fluctuation--Dissipation--Response Inequalities}

\author{Jie Gu}
\affiliation{Chengdu Academy of Educational Sciences, Chengdu 610036, China}
\email{jiegu1989@gmail.com}

\date{\today}

\begin{abstract}
We derive spectral fluctuation--dissipation--response inequalities for finite-state Markov jump processes.
By comparing the causal susceptibility to its passive equilibrium reference, we establish frequency-resolved and frequency-integrated inequalities that bound their mismatch in terms of the steady-state entropy production rate, probe variance, short-time perturbation diffusion, and reversible relaxation timescales. Our bounds exactly recover the standard fluctuation--dissipation theorem at equilibrium and apply directly to measurable causal susceptibilities, providing experimentally testable thermodynamic limits on FDT breakdown in driven steady states.
\end{abstract}

\maketitle

%\Large

%============================================================

%==============================================================
%==============================================================
\paragraph{Introduction.---}
\label{sec:intro}
%==============================================================

The fluctuation--dissipation theorem (FDT) is a cornerstone of statistical mechanics, providing a profound link between a system's spontaneous equilibrium fluctuations and its linear response to weak external perturbations \cite{Kubo1966}. In equilibrium, this principle offers a powerful advantage: one can perfectly predict how a system will react to a driving field using only passive measurements of its background noise. However, for systems operating far from equilibrium—such as active colloids, molecular motors, and living cellular networks—detailed balance is broken by continuous energy consumption, and the standard FDT generically fails. 

Despite this breakdown, estimating response out of equilibrium remains a central experimental challenge. In many physical and biological scenarios, stationary fluctuations and lock-in responses are readily measurable, whereas microscopic thermodynamic forces, internal cyclic currents, and the total entropy production rate remain deeply hidden from the observer. Consequently, a pragmatic experimental approach is to still evaluate the equilibrium-style fluctuation--dissipation predictor, $\chi_{\text{eq}}(\omega)$, and compare it against the actively measured causal susceptibility, $\chi(\omega)$. The discrepancy between them, $\Delta\chi(\omega) \equiv \chi(\omega) - \chi_{\text{eq}}(\omega)$, is not merely an estimation error; it is a fundamental, frequency-resolved macroscopic signature of nonequilibrium activity. 

Existing theoretical frameworks brilliantly illuminate the origins of this breakdown, yet they often do not constrain $\Delta\chi(\omega)$ in an experimentally direct manner. Generalized FDTs express the response through entropy-related and dynamical-activity contributions within stochastic thermodynamics \cite{BaiesiMaesWynants2009,Seifert2012}, and for suitable observables one can sometimes recover an equilibrium-like form by redefining the conjugate fluctuation \cite{SpeckSeifert2006,ProstJoannyParrondo2009}. Harada--Sasa-type relations connect integrated FDT violations to dissipation in Langevin steady states \cite{HaradaSasa2005}, while Markov jump processes can display genuinely frequency-dependent violations, particularly at high frequencies \cite{Wang2018}. Concurrently, thermodynamic uncertainty relations and related response bounds show that, even far from equilibrium, fluctuations and response remain constrained by entropy production and dynamical activity \cite{BaratoSeifert2015,gingrich2016dissipation, DechantSasa2020,owen2020,owen2023,fernandesmartins2023,aslyamov2024a,aslyamov2024,zheng2025,liu2025, kwon2025fluctuation, van2025fundamental, liu2025response, bao2025}. Recent progress has extended this perspective into the frequency domain. Dechant showed that dissipation and relaxation leave characteristic signatures in frequency-resolved fluctuations \cite{Dechant2023PSD}, and later derived a finite-frequency fluctuation--response inequality for general Markovian dynamics \cite{Dechant2025FFRIneq}. Zheng and Lu obtained related finite-frequency bounds for steady-state Markov processes under time-dependent perturbations \cite{ZhengLu2026FFR}. In parallel, exact nonequilibrium fluctuation--response relations and macroscopic spectral theories have been established for Markov jump networks \cite{aslyamov2024a,aslyamov2024,Aslyamov2025MacroscopicFR}. What remains missing amidst these advances, however, is a universal spectral bound on the exact quantity directly confronted in active measurement: the deviation of the physical causal susceptibility from its passive equilibrium predictor.

In this Letter, we bridge this gap for finite-state continuous-time Markov jump processes governed by local detailed balance. Rather than attempting to exactly reconstruct the complex nonequilibrium response, we ask a more robust question: how large can the equilibrium-style estimation error be, and what physical principles bound it? We derive a family of spectral fluctuation--dissipation--response inequalities (FDRIs) that rigorously bound this  mismatch. Formulated both pointwise in frequency and integrated over all frequencies, these bounds constrain the squared FDT violation using the steady-state entropy production rate, the variance of the readout observable, the short-time diffusion scale of the perturbation, and the system's reversible relaxation timescales. By locking the breakdown of the FDT beneath a strict thermodynamic and kinetic ceiling, our results transform an abstract nonequilibrium feature into a directly testable constraint for small-signal experiments.

%==============================================================
\paragraph{Setup and main results.---}
\label{sec:setup}

We consider a continuous--time Markov jump process on a finite state space $\Omega=\{1,\dots,n\}$ with rate matrix $W$.
Off--diagonal entries $W_{ij}\ge 0$ ($i\neq j$) are transition rates from $j$ to $i$, and $\sum_i W_{ij}=0$.
Irreducibility guarantees a unique stationary distribution $\pi$ with $\pi_i>0$.
We also write the stationary edge fluxes and currents as $a_{ij}\equiv \pi_j W_{ij}$, $J_{ij}\equiv a_{ij}-a_{ji}$, and $A_{ij}\equiv a_{ij}+a_{ji}$.

We focus on a state observable $r:\Omega\to\mathbb{R}$ with $\langle r\rangle_\pi \equiv \sum_i \pi_i r(i) = 0$.
We write $r(i)$ for its value on state $i$, and collect these values into the state-indexed column vector $r$.
We denote stationary averages by $\langle\cdot\rangle_\pi$ and use the inner product $\langle f,g\rangle_\pi=\sum_i \pi_i f(i)g(i)$.
Its autocorrelation and power spectrum are
\begin{equation*}
C_{rr}(t) = \langle r(t)r(0)\rangle_\pi,
\qquad
\mathcal{S}_{rr}(\omega) = \int_{-\infty}^{\infty} dt\,e^{i\omega t} C_{rr}(t).
\end{equation*}
For a perturbation conjugate to a state observable $b:\Omega\to\mathbb{R}$, with centered version $\delta b=b-\langle b\rangle_\pi$, we also use the cross--correlation
\begin{equation*}
C_{rb}(t) = \langle r(t)\delta b(0)\rangle_\pi.
\end{equation*}

A weak field $\varepsilon h(t)$ couples to $b$ through the local-detailed-balance--preserving tilting \cite{Maes2010ResponseFormula}
\begin{equation}
W^{\varepsilon}_{ij}(t)
=
W_{ij}\exp\!\Bigl[\varepsilon \beta h(t)\bigl(\eta \,b(j)-\zeta \,b(i)\bigr)\Bigr],
\qquad i\neq j.
\label{eq:rates-perturb}
\end{equation}
with $\eta +\zeta =1$ and $W^{\varepsilon}_{ii}(t)=-\sum_{k\neq i}W^{\varepsilon}_{ki}(t)$.
The physical linear-response kernel is defined by
\begin{equation*}
\delta\langle r(t)\rangle
= \varepsilon\int_{-\infty}^{\infty} dt'\,\chi(t-t')h(t') + O(\varepsilon^2),
\qquad
\chi(t<0)=0,
\end{equation*}
and its one-sided Fourier/Laplace transform
\begin{equation*}
\chi(\omega)
\equiv \int_0^{\infty} dt\,e^{i\omega t}\chi(t)
\end{equation*}
is the measurable causal susceptibility.
From passive steady-state fluctuations of the same pair $(r,b)$, we define the equilibrium-style FDT reference
\begin{equation*}
\chi_{\mathrm{eq}}(\omega)
\equiv\beta\int_0^{\infty} dt\,e^{i\omega t}\,\partial_t C_{rb}(t).
\end{equation*}
This $\chi_{\mathrm{eq}}(\omega)$ is directly measurable from stationary correlations of $r$ and $b$. Under detailed balance it coincides with the actual susceptibility, $\chi(\omega)=\chi_{\mathrm{eq}}(\omega)$; away from equilibrium it is only a passive reference predictor.

To state the bounds we also use the observable generator $W^{\mathsf T}$,
\begin{equation*}
(W^{\mathsf T}f)(i)=\sum_{j\neq i}W_{ji}[f(j)-f(i)],
\end{equation*}
its adjoint in the $\pi$--inner product,
\begin{equation*}
W^\dagger\equiv\Pi^{-1}W\Pi,
\qquad
\Pi=\mathrm{diag}(\pi),
\end{equation*}
and the symmetric and antisymmetric parts $W_{\mathrm{sym}}\equiv(W^{\mathsf T}+W^\dagger)/2$ and $W_{\mathrm{asym}}\equiv(W^{\mathsf T}-W^\dagger)/2$.
The spectral gap $\lambda>0$ is the smallest nonzero eigenvalue of $-W_{\mathrm{sym}}$.
As shown in the End Matter, the mismatch admits the causal representation $\Delta\chi(\omega)=\int_0^\infty dt\,e^{i\omega t}C_{rv}(t)$ with $C_{rv}(t)\equiv \langle r(t)v(0)\rangle_\pi$ and $v\equiv 2\zeta\beta W_{\mathrm{asym}}b$.

The quantity of central interest is the mismatch $\chi(\omega)-\chi_{\mathrm{eq}}(\omega)$. 
It is a causal, analytic, mode-resolved measure of the failure of passive fluctuations to encode the full linear response out of equilibrium, and singles out the part of the irreversible steady current sector that is injected by the perturbation coordinate.
 The low-frequency regime probes the static nonequilibrium defect and its higher temporal moments. The high-frequency regime is controlled by equal-time overlaps and decays at least as $1/|\omega|$. See the Supplemental Material (SM) for more details of its properties.

Our main results are, as proven in the End Matter, two causal fluctuation--dissipation--response inequalities (FDRIs).
The first one is a frequency-resolved FDRI, given by
\begin{equation}
\bigl|\chi(\omega)-\chi_{\mathrm{eq}}(\omega)\bigr|^2
\le  \frac{4\zeta^2\beta^2 D_b}{\pi_{\min} \lambda^2} \, \mathrm{Var}(r)\,\sigma ,
\label{eq:frequency-wise-FDRI}
\end{equation}
where $\sigma$ is the steady--state entropy production rate, $\mathrm{Var}(r) \equiv \langle r^2 \rangle_\pi$ is the equal-time variance of $r$,
$D_b\equiv\tfrac12\sum_{i,j}\pi_j W_{ij}[b(i)-b(j)]^2$ is the short--time diffusion coefficient of $b$, and $\pi_{\min}=\min_i \pi_i$ the minimal stationary probability.

For the frequency-integrated inequality, we have 
\begin{equation}
\begin{aligned}
\frac{1}{2\pi}\int_{-\infty}^{\infty}\! & d\omega\,
\bigl|\chi(\omega)-\chi_{\mathrm{eq}}(\omega)\bigr|^2
\;\le\; \\
&\min\!\left\{
\frac{2\zeta^2 \beta^2}{\pi_{\min}\lambda}\,D_b\,\mathrm{Var}(r),
\frac{4\zeta^2 \beta^2}{\pi_{\min}}\,D_b\,\|\mathcal{S}_{rr}\|_{\infty}
\right\}\sigma,	
\end{aligned}
\label{eq:integrated-FDRIs}
\end{equation}
where $\|\mathcal{S}_{rr}\|_{\infty}\equiv\sup_{\omega\in\mathbb{R}}\mathcal{S}_{rr}(\omega)$.
A sharper version of both bounds replaces $4D_b/\pi_{\min}$ by a local coupling prefactor $\Gamma_b$ defined in the End Matter.

%%==============================================================
\paragraph{Physical interpretation and implications.---}
\label{sec:interpretation}

The physical message is that dissipation limits not only current precision \cite{BaratoSeifert2015,gingrich2016dissipation}, the speed of stochastic evolution \cite{shiraishi2018,falasco2020dissipation}, and cross-correlation asymmetry \cite{ohga2023thermodynamic,liang2023thermodynamic,shiraishi2023entropy,van2024dissipation,gu2024thermodynamic}, but also the failure of passive fluctuations to predict the full causal response out of equilibrium. Indeed,
$
\Delta\chi(\omega)
=\int_{0}^{\infty}dt\,e^{i\omega t}\,\langle r(t)v(0)\rangle_{\pi},
$ with $
v=2\zeta\beta W_{\mathrm{asym}}b,
$
so the mismatch isolates the current-carrying, time-antisymmetric sector of the dynamics. It therefore vanishes at detailed balance, where $W_{\mathrm{asym}}=0$, recovering the FDT.

The bounds show that entropy production alone does not fix the observable violation; it must be filtered through the kinetics of the chosen perturbation and readout. Here $\sigma$ controls the size of the irreversible sector, $D_b$ quantifies how strongly nonequilibrium currents couple into the perturbation channel, $\mathrm{Var}(r)$ sets the observable scale of the readout, and $\lambda^{-1}$---or $\|\mathcal{S}_{rr}\|_{\infty}$ in the gap-free integrated form---sets the timescale over which the defect can accumulate before reversible relaxation suppresses it. This also explains the different powers of $\lambda$: the frequency-resolved bound integrates the decaying kernel before squaring, giving $\lambda^{-2}$, whereas the integrated bound squares first and therefore carries only $\lambda^{-1}$.

Near equilibrium, if thermodynamic forces are scaled as $O(\epsilon)$, then $J_{ij}=O(\epsilon)$, $\sigma=O(\epsilon^2)$, and $v=O(\epsilon)$, so that $|\Delta\chi(\omega)|^2=O(\sigma)$ both pointwise and after frequency integration. The FDT breakdown is therefore perturbatively weak---at most $O(\sqrt{\sigma})$ in amplitude and $O(\sigma)$ in integrated spectral weight---and the inequalities are asymptotically sharp in scaling.

The dependence on $\zeta$ is kinetic rather than thermodynamic. All perturbations in the local-detailed-balance family
$
W^\varepsilon_{ij}(t)=W_{ij}\exp\!\big[\varepsilon\beta h(t)\big(\eta b(j)-\zeta b(i)\big)\big]
$
have the same antisymmetric log-ratio, and hence the same excess entropy flux; changing $\zeta$ only redistributes the time-symmetric activity, i.e., the frenetic contribution \cite{BaiesiMaesWynants2009,Maes2010ResponseFormula,Seifert2012}. Thus $\zeta$ does not measure the distance from equilibrium, but how strongly the chosen perturbation protocol projects the antisymmetric current sector onto the readout channel, which also explains the $\zeta^2$ dependence of the bounds.

\paragraph{Comparison and connection with prior work.---}
\label{sec:connections}

Generalized fluctuation--dissipation relations express linear response as correlations with conjugate observables and separate entropic and frenetic contributions \cite{BaiesiMaesWynants2009,Seifert2012}. Recent finite-frequency inequalities instead bound the \emph{full} response--fluctuation form by spectral positivity or related Cauchy--Schwarz arguments \cite{Dechant2025FFRIneq,ZhengLu2026FFR}, while macroscopic fluctuation theory can reconstruct effective drift and diffusion directly from spectral data \cite{Aslyamov2025MacroscopicFR}. By contrast, our result acts on the defect $\Delta\chi(\omega)=\chi(\omega)-\chi_{\mathrm{eq}}(\omega)$, namely the component of the causal response transverse to the detailed-balance manifold. Since $\Delta\chi(\omega)$ vanishes at detailed balance and is sourced only by the antisymmetric current sector through $v(i)=\beta\zeta\sum_j J_{ij}[b(i)-b(j)]/\pi_i$, the theorem is best viewed as a stability estimate around equilibrium rather than a generic bound on response amplitude.

Combined with $\sum_{i,j}J_{ij}^2/A_{ij}\le \sigma$ and reversible semigroup decay, the bound factorizes into three ingredients: dissipation, coupling geometry, and mixing. What is controlled here is therefore not the ability of a system to respond, but the failure of passive fluctuations to encode the full causal response out of equilibrium. The mismatch survives only when steady currents cross edges on which the perturbation coordinate varies and when the readout overlaps with slowly relaxing modes. In this sense the result is closer in spirit to Harada--Sasa-type relations \cite{HaradaSasa2005}, which also isolate the genuinely nonequilibrium part of the response, but here the statement applies to finite-state jump processes directly at finite frequency for the one-sided causal susceptibility. The integrated bound may moreover be viewed as a one-sided causal counterpart of bilateral spectral bounds \cite{Dechant2025FFRIneq}.

%==============================================================
\paragraph{Experimental Relevance.---}
\label{sec:experimental}
%==============================================================

The quantities on the left-hand side are  directly measurable. The susceptibility $\chi(\omega)$ is obtained by a weak sinusoidal modulation of the field conjugate to $b$ and lock-in (or equivalent small-signal) measurement of the response of $r$. The reference $\chi_{\mathrm{eq}}$ is constructed from passive steady-state fluctuations of $r$ and $b$ alone. Thus the violation itself can be measured without first inferring a Markov-state model.

Among those prefactors on the right-hand side, $\mathrm{Var}(r)$ is simply the equal-time variance of the readout and $\|\mathcal{S}_{rr}\|_{\infty}$ is the maximal passive noise level, both directly obtainable from trajectory data. Likewise, $D_b$ is the short-time growth rate of the variance of $b$ along the trajectory. The spectral gap $\lambda$ is more challenging: the slowest observed decay rate of a single correlation function generally yields at best an {upper} bound on $\lambda$; obtaining a rigorous lower bound requires either full model inference or additional structural information. For this reason the second integrated estimate in Eq.~\eqref{eq:integrated-FDRIs}, which avoids $\lambda$ in favor of $\|\mathcal{S}_{rr}\|_{\infty}$, is usually the more robust data-analysis statement when only passive fluctuations and short-time increments are accurately available. If a discrete-state model is reconstructed, then $\pi_{\min}$, $\sigma$, and the sharper prefactor $\Gamma_b$ can also be computed from the inferred generator.

Optically trapped colloids provide a clean nonequilibrium testbed: multiwell potentials produce discrete hopping, and weak perturbations are introduced by small trap adjustments. Modified fluctuation--dissipation relations and work/dissipation statistics have already been measured in such systems \cite{GomezSolano2009PRL,Jop2008EPL}. Related platforms extend this strategy: single-electron boxes enable trajectory-level entropy-production measurements \cite{Koski2013NatPhys}, single F$_1$-ATPase allows load-controlled susceptibility tests \cite{Toyabe2010PRL}, and ion-channel or smFRET trajectories can be analyzed with Markov/HMM methods to infer the relevant prefactors \cite{Oikonomou2024CommsChem,Roy2008NatMethods}. In the latter cases, small ligand or chemical-potential shifts can directly probe $\chi_{\mathrm{eq}}$ and $\chi$.

%==================================

\paragraph{Example: uniform unicyclic network.---}
As a solvable benchmark, consider an $N$-state ring with uniform forward and backward rates
$
W_{i+1,i}=k e^{F/2},\qquad
W_{i-1,i}=k e^{-F/2},
$
with $i$ understood modulo $N$. The stationary state is uniform, $\pi_i=1/N$. For a Fourier mode $q=2\pi m/N$, define
\[
c_q(i)=\sqrt{2}\cos(qi),\qquad
s_q(i)=\sqrt{2}\sin(qi).
\]
These span an invariant two-dimensional subspace,
\[
\begin{aligned}
&W_{\mathrm{sym}}c_q=-\mu_q c_q,\quad
W_{\mathrm{sym}}s_q=-\mu_q s_q,\\
&W_{\mathrm{asym}}c_q=-\Omega_q s_q,\quad
W_{\mathrm{asym}}s_q=\Omega_q c_q,	
\end{aligned}
\]
with
$
\mu_q=2k\cosh(F/2)\bigl(1-\cos q\bigr)$ and $
\Omega_q=2k\sinh(F/2)\sin q.
$

Choosing $b=c_q$ and $r=s_q$ gives
$
v=2\zeta\beta W_{\mathrm{asym}}b=-2\zeta\beta\Omega_q s_q,
$
so that the causal mismatch can be evaluated exactly (see SM):
\[
\Delta\chi_q(\omega)
=
-2\zeta\beta\Omega_q\,
\frac{\mu_q-i\omega}{(\mu_q-i\omega)^2+\Omega_q^2}.
\]
The nonequilibrium defect is therefore governed by a simple competition between reversible decay, set by $\mu_q$, and phase rotation, set by $\Omega_q$.

For the slowest mode $q=2\pi/N$, one has $\lambda=\mu_q$, and the frequency-resolved bound in Eq.~\eqref{eq:frequency-wise-FDRI} is approached with ratio
\[
\frac{|\Delta\chi_q(0)|}{\sqrt{\mathrm{Var}(r)\langle v^2\rangle_\pi}/\lambda}
=
\frac{1}{1+(\Omega_q/\mu_q)^2},
\]
where $
{\Omega_q}/{\mu_q}
=
\tanh(F/2)\cot(q/2).
$
Likewise, the gap-based integrated step behind Eq.~\eqref{eq:integrated-FDRIs} has exact ratio
\[
\frac{
\frac{1}{2\pi}\int d\omega\,|\Delta\chi_q(\omega)|^2
}{
\mathrm{Var}(r)\langle v^2\rangle_\pi/(2\lambda)
}
=
\frac12\!\left[
1+\frac{1}{1+(\Omega_q/\mu_q)^2}
\right].
\]
Thus both bounds become near tight when $\Omega_q/\mu_q\ll1$, i.e.\ when decay dominates phase rotation on the selected mode.

The thermodynamic closure is also explicit:
\[
\frac{\sum_{i,j}J_{ij}^2/A_{ij}}{\sigma}
=
\frac{2\tanh(F/2)}{F}
=
1-\frac{F^2}{12}+O(F^4).
\]
The ring therefore makes the near-saturation mechanism transparent: mode purity keeps $r$ and $v$ in the same invariant subspace, weak phase rotation preserves coherence in the one-sided transform, and weak driving renders the current--traffic step asymptotically tight.

%==============================================================

\paragraph{Example: ATP-driven push--pull phosphorylation switch.---}
A ubiquitous biochemical motif is the reversible covalent modification cycle, e.g.\ phosphorylation--dephosphorylation by a kinase and a phosphatase, often described as a ``push--pull'' system \cite{GoldbeterKoshland1981}.
Such cycles are routinely probed by periodic inputs and analyzed in the frequency domain \cite{SzemereRotsteinVentura2021}.
Here we use a minimal multicyclic Markov jump model to illustrate the causal FDRIs.

We coarse-grain a single substrate molecule into four states:
$0 \equiv X$ ({unphosphorylated}), $1\equiv XK$ ({kinase-bound}), $2\equiv X^\ast$ ({phosphorylated/active}), and $3\equiv X^\ast P$ ({phosphatase-bound}), 
with transitions
$X\leftrightarrow XK\leftrightarrow X^\ast\leftrightarrow X^\ast P\leftrightarrow X$
and an additional {bypass} edge $X^\ast\leftrightarrow X$ representing spontaneous dephosphorylation.
This adds a chord to the enzymatic cycle, producing two independent cycles.

The rates satisfy local detailed balance edgewise,
$\ln({W_{ij}}/{W_{ji}})
=\beta\bigl(E_j-E_i+w_{i\leftarrow j}\bigr)$, $w_{j\leftarrow i}=-w_{i\leftarrow j}$,
with dimensionless state energies $E_i$ and a nonequilibrium chemical work term $w_{i\leftarrow j}$.
Only the phosphorylation step $1\leftrightarrow 2$ is driven: we set $w_{2\leftarrow 1}=\Delta\mu$ and $w_{1\leftarrow 2}=-\Delta\mu$,
interpreting $\Delta\mu$ as an ATP chemical potential drop that biases phosphorylation.
All other edges have $w=0$.
The parameter $\Delta\mu$ controls dissipation: $\Delta\mu=0$ gives detailed balance, while $\Delta\mu>0$ yields a NESS with positive entropy production rate.

\begin{figure}[tb!]
    \centering
    \includegraphics[width=\linewidth]{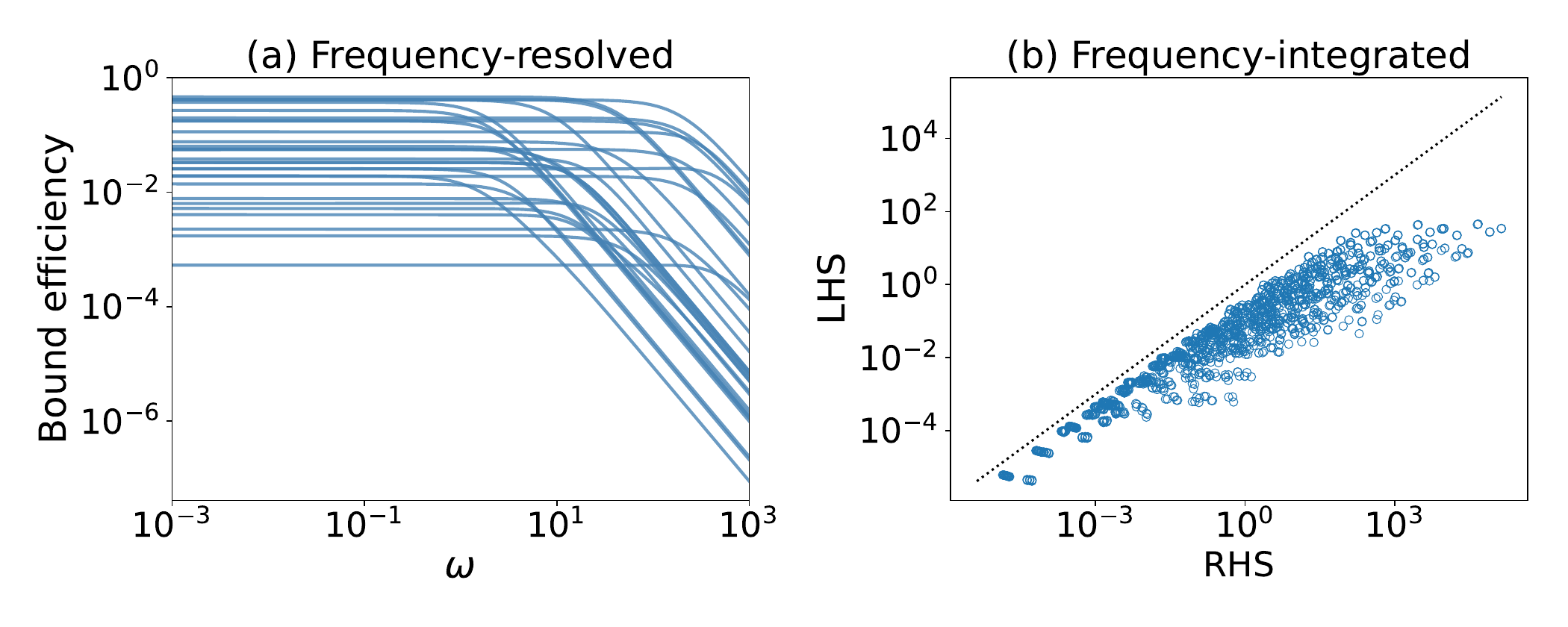}
\caption{
Bound efficiency of the causal inequality in the four-state push--pull network. (a) Frequency-resolved bound efficiency, defined as $\mathrm{LHS}/\mathrm{RHS}$, as a function of $\omega$ for 30 parameter combinations selected at random from the scanned set. (b) Frequency-integrated $\mathrm{LHS}$ plotted against the corresponding integrated $\mathrm{RHS}$ for the full scan of $1728$ parameter sets; the dotted line marks saturation of the bound. 
The state energies are $E=(0,0,E_2,E_3)$ with $E_2\in\{0.8,1.3,1.8\}$ and $E_3\in\{0,0.4,0.8\}$, the inverse temperature is $\beta\in\{0.4,1.333\ldots,2.267\ldots,3.2\}$, the driving affinity is $\Delta\mu\in\{0.5,2.0,3.5,5.0\}$, the ring rate $k_{\mathrm{ring}}$ takes four logarithmically spaced values from $1$ to $10^{2}$, and the bypass rate $k_{\mathrm{bypass}}$ takes three logarithmically spaced values from $10^{-3}$ to $1$.
}
    \label{fig:pushpull}
\end{figure}

We perturb the system by a weak ``energy-like'' field $h(t)$ that shifts the free energy of phosphorylated forms.
Concretely, we take the conjugate state observable $b(0)=b(1)=0, b(2)=b(3)=1$,
and use the local-detailed-balance tilting protocol \eqref{eq:rates-perturb} with symmetric splitting $\eta=\zeta=1/2$.
We use the experimentally accessible two-level reporter
$
r_{\rm br}=(0,1,1,0),
$
which can be interpreted as a fluorescence readout that is high on the kinase-side branch \(\{XK,X^\ast\}\) and low on \(\{X,X^\ast P\}\).

Figure \ref{fig:pushpull}(a) reports the frequency-resolved bound efficiency, defined as the ratio of the left- and right-hand sides of Eq.~\eqref{eq:frequency-wise-FDRI}, for \(30\) random parameter combinations.
All curves remain strictly below unity throughout the scanned frequency window, as required by the theorem.
The efficiency is typically largest at low and intermediate frequencies, where the nonequilibrium defect is most visible, and decreases at large \(\omega\), consistent with the generic high-frequency decay of the causal mismatch.
The integrated test is summarized in Fig. \ref{fig:pushpull}(b), where the left-hand side of Eq. \eqref{eq:integrated-FDRIs} is plotted against the corresponding bound.
Every point lies below the diagonal saturation line, confirming the integrated causal FDRI across a broad nonequilibrium regime.
The spread below the diagonal shows that the bound is universal rather than generically tight, while the points closest to the diagonal demonstrate that sizable causal FDT violations can still arise in a biochemically interpretable push--pull network without exceeding the thermodynamic ceiling imposed by the FDRIs.
This example therefore makes concrete the central message of the paper: ATP consumption can produce a measurable, frequency-dependent departure from the passive equilibrium-style response prediction, yet the size of that departure remains quantitatively limited by entropy production and relaxation-scale constraints.

%==============================================================
\paragraph{Conclusions and outlook.---}
\label{sec:conclusion}
%==============================================================

We have derived a family of fluctuation--dissipation--response inequalities for finite--state Markov jump processes with local detailed balance.
These inequalities bound the deviation of the {causal susceptibility} $\chi(\omega)$ from its equilibrium FDT reference $\chi_{\mathrm{eq}}(\omega)$ in terms of entropy production, the reversible spectral gap, and static or short--time fluctuation scales associated with the probe and perturbation observables. The inequalities act directly on the one-sided susceptibility measured in experiments.

We have focused on energy-like perturbations consistent with local detailed balance, as these provide the most natural experimental probes of state energetics. For more general kinetic, or nonpotential, perturbations, however, the response mismatch can acquire additional frenetic contributions and therefore need not be constrained by entropy production alone \cite{BaiesiMaesWynants2009,Maes2010ResponseFormula}. Several extensions are immediate. An important direction is to continuous-space Langevin dynamics, where spectral gaps and relaxation bounds can be formulated in terms of Dirichlet forms, Poincar\'e and log-Sobolev inequalities \cite{BakryGentilLedoux2014,Pavliotis2014,Villani2009Hypocoercivity}. It would also be interesting to derive sharper frequency-resolved bounds by isolating the dominant slow modes, for example through spectral decompositions of $W_{\mathrm{sym}}$. A further possibility is to combine our framework with information-theoretic measures of irreversibility, such as relative entropy or Kullback--Leibler divergences between forward and backward path ensembles, in order to obtain refined bounds on response \cite{KawaiParrondoVanDenBroeck2007,RoldanParrondo2012}. Finally, systematic experimental tests in platforms where both passive fluctuations and active small-signal response can be measured with high precision---including driven colloids, single-electron devices, and single-molecule motors---could clarify how closely real nonequilibrium systems approach saturation of the causal FDRIs \cite{GomezSolano2009PRL,Koski2013NatPhys,Toyabe2010PRL}.

\medskip

\begin{acknowledgements}
	We would like to thank Shiling Liang for helpful discussions and valuable comments on an earlier draft of this manuscript
\end{acknowledgements}

\bibliographystyle{apsrev4-1}
\bibliography{ref}

%\clearpage

\section{End Matter}

Recall the stationary fluxes $a_{ij}=\pi_j W_{ij}$, net currents
$J_{ij}=a_{ij}-a_{ji}, J_{ji}=-J_{ij}$, symmetric activities $A_{ij}=a_{ij}+a_{ji}$, and thermodynamic forces
$F_{ij} = \ln {a_{ij}}/{a_{ji}}, F_{ji}=-F_{ij}$. The steady--state entropy production rate is
$\sigma = \frac{1}{2}\sum_{i,j} J_{ij}F_{ij}\ge 0$ with equality if and only if detailed balance holds.

For the energy--like perturbations considered in the main text, the weak field $\varepsilon h(t)$ couples to the state observable $b$ by a local-detailed-balance--preserving tilting of the jump rates ($i\neq j$) \cite{Maes2010ResponseFormula},
\begin{equation*}
W^{\varepsilon}_{ij}(t)
=
W_{ij}\exp\!\Bigl[\varepsilon \beta h(t)\bigl(\eta \,b(j)-\zeta \,b(i)\bigr)\Bigr],
\quad \eta +\zeta =1.
\end{equation*}
with $W^{\varepsilon}_{ii}(t)=-\sum_{k\neq i}W^{\varepsilon}_{ki}(t)$.
Let $(W^{\varepsilon}(t))^{\mathsf T}$ be the corresponding generator acting on observables as $((W^{\varepsilon}(t))^{\mathsf T}f)(i)=\sum_{j\neq i}W^\varepsilon_{ji}(t)\,[f(j)-f(i)]$, and define the causal linear-response kernel $\chi$ by the first-order change in the stationary expectation,
$\delta\langle r(t)\rangle
=\varepsilon\int_{-\infty}^{\infty}\!dt'\,\chi(t-t')\,h(t')+O(\varepsilon^2)$,
with $\chi(t)=0$ for $t<0$ and $X_0\sim\pi$ for the unperturbed dynamics.
Writing
\begin{equation*}
W^{(1)}_{ij}\equiv\left.\partial_\varepsilon W^\varepsilon_{ij}(t)\right|_{\varepsilon=0,\;h=1}
=\beta (\eta \,b(j)-\zeta \,b(i))\,W_{ij}\qquad (i\neq j),
\end{equation*}
and using the Duhamel (variation-of-constants) expansion of the perturbed semigroup around the stationary state, one obtains the Agarwal representation \cite{Agarwal1972FDT,SeifertSpeck2010EPL}
\begin{equation*}
\chi(t)=\Theta(t)\langle e^{tW^{\mathsf T}}r,\psi\rangle_\pi=\Theta(t)\langle r(t)\,\psi(0)\rangle_\pi,
\end{equation*}
where $\Theta$ is the Heaviside step function, $\psi\equiv\Pi^{-1}W^{(1)}\pi$, and $\Pi \equiv \mathrm{diag}(\pi)$.
Evaluating $\psi$ explicitly in terms of the unperturbed steady-state fluxes, currents and activities gives the closed state-observable form
\begin{equation}
\begin{aligned}
\psi(i)
& =
\beta\big[\zeta (W^{\mathsf T}b)(i)+\eta (W^\dagger b)(i)\big]
\\
&=\beta \big[W_{\mathrm{sym}}b(i)-(\eta-\zeta)W_{\mathrm{asym}}b(i)\big],
\end{aligned}
\label{eq:psi-explicit}
\end{equation}
with $W_{\mathrm{sym}}=(W^{\mathsf T}+W^\dagger)/2$ and $W_{\mathrm{asym}}=(W^{\mathsf T}-W^\dagger)/2$.

The measurable susceptibility is the one-sided transform
\begin{equation*}
\chi(\omega)
\equiv\int_0^\infty dt\,e^{i\omega t}\chi(t)
=\int_0^\infty dt\,e^{i\omega t}\,C_{r\psi}(t),
\end{equation*}
with
\begin{equation*}
	C_{r\psi}(t)\equiv\langle r(t)\psi(0)\rangle_\pi.
\end{equation*}
For the equilibrium reference we choose $\psi_{\mathrm{eq}}\equiv\beta\,W^\dagger b$, which depends only on $b$ and the unperturbed generator, and define
\begin{equation*}
\chi_{\mathrm{eq}}(t)\equiv\Theta(t)\langle r(t)\psi_{\mathrm{eq}}(0)\rangle_\pi,
\qquad
\chi_{\mathrm{eq}}(\omega)\equiv\int_0^\infty dt\,e^{i\omega t}\chi_{\mathrm{eq}}(t).
\end{equation*}
Since
\begin{equation*}
\partial_t C_{rb}(t)
=\partial_t\langle e^{tW^{\mathsf T}}r,\delta b\rangle_\pi
=\langle e^{tW^{\mathsf T}}r,W^\dagger b\rangle_\pi,
\end{equation*}
we have the passive representation
\begin{equation*}
\chi_{\mathrm{eq}}(t)=\beta\Theta(t)\,\partial_t C_{rb}(t).
\end{equation*}
When detailed balance holds, $W^\dagger=W^{\mathsf T}$ and therefore $\psi=\psi_{\mathrm{eq}}$, so $\chi(\omega)=\chi_{\mathrm{eq}}(\omega)$.

Define the violation observable
\begin{equation*}
v  \equiv \psi-\psi_{\mathrm{eq}},
\end{equation*}
which is automatically centered, $\langle v \rangle_\pi=0$.
Then
\begin{equation*}
C_{rv}(t)
= \langle r(t)v(0)\rangle_\pi
= C_{r\psi}(t)-C_{r\psi_{\mathrm{eq}}}(t),
\end{equation*}
and hence
\begin{equation*}
\Delta\chi(\omega)
\equiv\chi(\omega)-\chi_{\mathrm{eq}}(\omega)
= \int_{0}^{\infty} dt\,e^{i\omega t}C_{rv}(t).
\end{equation*}
Using Eq.~\eqref{eq:psi-explicit},
\begin{equation*}
v =\psi-\psi_{\mathrm{eq}}=\beta\zeta\,(W^{\mathsf T}-W^\dagger)b=2\zeta\beta\,W_{\mathrm{asym}}b.
\end{equation*}
In current form this becomes
\begin{equation}
v(i)=\frac{\beta\zeta}{\pi_i}\sum_j J_{ij}\,[b(i)-b(j)],
\label{eq:A-chi-current}
\end{equation}

\paragraph{Derivation of the frequency-resolved FDRI.---}
\label{app:frequency-wise}
Let $W^{\mathsf T}$ be the generator acting on observables and $W_{\mathrm{sym}}$ its symmetrized version, with spectral gap $\lambda$.
By the min-max theorem,
\begin{equation*}
- \langle f, W_{\mathrm{sym}} f\rangle_\pi
\ge \lambda\langle f^2\rangle_\pi
\end{equation*}
for all mean--zero observables $f$.
Set $g_t = e^{tW^{\mathsf T}}r$.
Since $\langle r\rangle_\pi=0$, also $\langle g_t\rangle_\pi=0$, and therefore
\begin{equation*}
\frac{d}{dt}\|g_t\|_\pi^2
= 2 \langle g_t, W_{\mathrm{sym}}g_t\rangle_\pi
\le -2\lambda\|g_t\|_\pi^2.
\end{equation*}
Gr\"onwall's lemma yields
\begin{equation*}
\|g_t\|_\pi \le e^{-\lambda t}\|r\|_\pi =\sqrt{\mathrm{Var}(r)}\,e^{-\lambda t}
\qquad (t\ge 0).
\end{equation*}
Hence, for $t\ge 0$,
\begin{equation}
|C_{rv}(t)|
= |\langle e^{tW^{\mathsf T}}r,v\rangle_\pi|
\le \|g_t\|_\pi\,\|v\|_\pi
\le e^{-\lambda t}\sqrt{\mathrm{Var}(r)\,\langle v^2\rangle_\pi}.
\label{eq:Crv-decay}
\end{equation}
Integrating the absolute value of the one-sided transform gives
\begin{equation*}
|\Delta\chi(\omega)|
\le \int_0^\infty dt\,|C_{rv}(t)|
\le \frac{1}{\lambda}\sqrt{\mathrm{Var}(r)\,\langle v^2\rangle_\pi}.
\end{equation*}
Therefore
\begin{equation}
|\Delta\chi(\omega)|^2
\le \frac{1}{\lambda^2}\,\mathrm{Var}(r)\,\langle v^2\rangle_\pi.
\label{eq:pointwise-pre}
\end{equation}

It remains to bound $\langle v^2\rangle_\pi$ in thermodynamic terms.
From Eq.~\eqref{eq:A-chi-current},
\begin{equation*}
\langle v^2\rangle_\pi
= \sum_i \pi_i v(i)^2
= \beta^2\zeta^2\sum_i\frac{1}{\pi_i}
\left(\sum_j J_{ij}\Delta b_{ij}\right)^2,
\end{equation*}
with $\Delta b_{ij}\equiv b(i)-b(j)$.
For each $i$, apply Cauchy--Schwarz to the sum over $j$ with weights $A_{ij}$:
\begin{equation*}
\left(\sum_j J_{ij}\Delta b_{ij}\right)^2
\le
\left(\sum_j \frac{J_{ij}^2}{A_{ij}}\right)
\left(\sum_j A_{ij}\Delta b_{ij}^2\right).
\end{equation*}
Hence
\begin{equation*}
\langle v^2\rangle_\pi
\le \beta^2 \zeta^2 \sum_i\frac{1}{\pi_i}
\left(\sum_j \frac{J_{ij}^2}{A_{ij}}\right)
\left(\sum_j A_{ij}\Delta b_{ij}^2\right).
\end{equation*}
Introducing
\begin{equation*}
\Gamma_b
\equiv \max_{i\in\Omega}
\frac{1}{\pi_i}\sum_j A_{ij}[b(i)-b(j)]^2,
\end{equation*}
and using
$\sum_{i,j}\frac{J_{ij}^2}{A_{ij}}
\le \sigma
$ \cite{shiraishi2018,Shiraishi2021a,Dechant2022a}, we obtain
\begin{equation}
\langle v^2\rangle_\pi
\le \beta^2\zeta^2\,\Gamma_b\,\sigma.
\label{eq:v2-Gamma}
\end{equation}
Now let us bound $\Gamma_b$ by coarse steady-state quantities.
Define the short--time diffusion coefficient of $b$ in the steady state as
\begin{equation*}
D_b
\equiv \frac{1}{2}\lim_{\Delta t\to 0}
\frac{1}{\Delta t}\,
\langle [b(X_{\Delta t})-b(X_0)]^2\rangle_\pi
= \frac{1}{2}\sum_{i,j} a_{ij}[b(i)-b(j)]^2,
\end{equation*}
where the last equality follows from the Poisson statistics of jumps.
Since $A_{ij}=a_{ij}+a_{ji}$,
\begin{equation*}
\sum_{i,j}A_{ij}[b(i)-b(j)]^2
= 2\sum_{i,j} a_{ij}[b(i)-b(j)]^2
= 4D_b.
\end{equation*}
Using $\pi_{\min}\equiv\min_i\pi_i$ we have
\begin{equation}
\begin{aligned}
\Gamma_b
= \max_i \frac{1}{\pi_i}\sum_j A_{ij}[b(i)-b(j)]^2
&\le \frac{1}{\pi_{\min}}
\sum_{i,j}A_{ij}[b(i)-b(j)]^2 \\
&= \frac{4D_b}{\pi_{\min}}.
\end{aligned}
\label{eq:Gamma-to-Db}
\end{equation}
Combining Eqs.~\eqref{eq:pointwise-pre}, \eqref{eq:v2-Gamma}, and \eqref{eq:Gamma-to-Db} yields the frequency-resolved FDRI~\eqref{eq:frequency-wise-FDRI}. The sharper prefactor version is obtained by stopping at Eq.~\eqref{eq:v2-Gamma}.

\paragraph{Derivation of the integrated FDRIs.---}
\label{app:integrated}
By Parseval's identity for the one-sided transform,
\begin{equation*}
\frac{1}{2\pi}\int_{-\infty}^{\infty}d\omega\,|\Delta\chi(\omega)|^2
= \int_0^\infty dt\,|C_{rv}(t)|^2.
\end{equation*}
Using Eq.~\eqref{eq:Crv-decay},
\begin{equation*}
\begin{aligned}
\int_0^\infty dt\,|C_{rv}(t)|^2
&\le \mathrm{Var}(r)\,\langle v^2\rangle_\pi\int_0^\infty dt\,e^{-2\lambda t}
\\
&= \frac{1}{2\lambda}\,\mathrm{Var}(r)\,\langle v^2\rangle_\pi.
\end{aligned}
\end{equation*}
Combining with Eqs.~\eqref{eq:v2-Gamma} and \eqref{eq:Gamma-to-Db} gives the first integrated estimate,
\begin{equation}
\frac{1}{2\pi}\int d\omega\,|\Delta\chi(\omega)|^2
\le \frac{2\beta^2\zeta^2D_b}{\pi_{\min}\lambda}\,\mathrm{Var}(r)\,\sigma.
\label{eq:int-gap-version}
\end{equation}

A second estimate avoids $\lambda$ at the price of using the passive spectrum. Define the bilateral transform of the same violation kernel,
\begin{equation*}
\Delta\mathcal{R}(\omega)\equiv\mathcal{S}_{rv}(\omega)=\int_{-\infty}^{\infty}dt\,e^{i\omega t}C_{rv}(t).
\end{equation*}
Since $\Theta(t)$ is a contraction on $L^2(\mathbb{R})$,
\begin{equation*}
\begin{aligned}
\frac{1}{2\pi}\int d\omega\,|\Delta\chi(\omega)|^2
&=\|\Theta C_{rv}\|_{L^2(\mathbb{R})}^2 \\
&\le \|C_{rv}\|_{L^2(\mathbb{R})}^2
= \frac{1}{2\pi}\int d\omega\,|\Delta\mathcal{R}(\omega)|^2.	
\end{aligned}
\end{equation*}
The $2\times 2$ spectral matrix
\begin{equation*}
\mathbb{S}(\omega) =
\begin{pmatrix}
\mathcal{S}_{rr}(\omega) & \mathcal{S}_{rv}(\omega)\\
\mathcal{S}_{vr}(\omega) & \mathcal{S}_{vv}(\omega)
\end{pmatrix}
\end{equation*}
is a power spectral density matrix and is thus Hermitian and nonnegative definite for every real $\omega$ \cite{Dechant2025FFRIneq}.
Therefore,
\begin{equation*}
|\Delta\mathcal{R}(\omega)|^2
=|\mathcal{S}_{rv}(\omega)|^2
\le \mathcal{S}_{rr}(\omega)\mathcal{S}_{vv}(\omega).
\end{equation*}
Integrating and using Wiener--Khinchin,
\begin{equation*}
\begin{aligned}
\frac{1}{2\pi}\int d\omega\,|\Delta\chi(\omega)|^2
&\le \frac{1}{2\pi}\int d\omega\,\mathcal{S}_{rr}(\omega)\mathcal{S}_{vv}(\omega)
\\
&\le \|\mathcal{S}_{rr}\|_\infty\,\frac{1}{2\pi}\int d\omega\,\mathcal{S}_{vv}(\omega)
\\
&= \|\mathcal{S}_{rr}\|_\infty\,\langle v^2\rangle_\pi.
\end{aligned}
\end{equation*}
Combining this with Eqs.~\eqref{eq:v2-Gamma} and \eqref{eq:Gamma-to-Db} yields the second integrated estimate,
\begin{equation}
\frac{1}{2\pi}\int d\omega\,|\Delta\chi(\omega)|^2
\le \frac{4\beta^2\zeta^2D_b}{\pi_{\min}}\,\|\mathcal{S}_{rr}\|_\infty\,\sigma.
\label{eq:int-sup-version}
\end{equation}
Taking the minimum of Eqs.~\eqref{eq:int-gap-version} and \eqref{eq:int-sup-version} gives Eq.~\eqref{eq:integrated-FDRIs}.

\clearpage
\onecolumngrid

\setcounter{page}{1}

\begin{center}
\textbf{Supplemental material for Spectral Fluctuation--Dissipation--Response Inequalities}

\vspace{0.5em}

Jie Gu
\end{center}

\vspace{1em}

This supplement collects structural properties of the causal fluctuation--dissipation mismatch
\begin{equation*}
\Delta\chi(\omega)\equiv \chi(\omega)-\chi_{\mathrm{eq}}(\omega)
=\int_0^\infty dt\,e^{i\omega t}C_{rv}(t),
\qquad
C_{rv}(t)=\langle r(t)v(0)\rangle_\pi,
\end{equation*}
with
\begin{equation*}
v\equiv \psi-\psi_{\mathrm{eq}}=2\zeta\beta\,W_{\mathrm{asym}}b,
\qquad
W_{\mathrm{asym}}\equiv \frac{W^{\mathsf T}-W^\dagger}{2}.
\end{equation*}
All notation is as in the main text. The state space is finite, the Markov process is irreducible, $r$ is a centered readout observable, and $b$ is the state observable coupled to the weak energy-like perturbation.

\section*{1. Basic representation and centeredness}

The mismatch is itself a causal susceptibility. Since $W_{\mathrm{asym}}$ is antisymmetric in the $\pi$-inner product and annihilates constants, one has
\begin{equation*}
\langle v\rangle_\pi
=
2\zeta\beta\,\langle 1,W_{\mathrm{asym}}b\rangle_\pi
=
-2\zeta\beta\,\langle W_{\mathrm{asym}}1,b\rangle_\pi
=0.
\end{equation*}
Thus both $r$ and $v$ belong to the centered subspace
\begin{equation*}
\mathcal H_0\equiv \bigl\{f:\Omega\to\mathbb R\ \big|\ \langle f\rangle_\pi=0\bigr\}.
\end{equation*}
Let $W^\dagger_\perp$ denote the restriction of $W^\dagger$ to $\mathcal H_0$. Its spectrum lies strictly in the open left half-plane. Using
\begin{equation*}
C_{rv}(t)=\langle r,e^{W^\dagger t}v\rangle_\pi,
\end{equation*}
one obtains the resolvent representation
\begin{equation}
\Delta\chi(\omega)
=
\bigl\langle r,\bigl(-W^\dagger_\perp-i\omega\bigr)^{-1}v\bigr\rangle_\pi.
\label{eq:SM-resolvent}
\end{equation}
Hence $\Delta\chi$ is the matrix element of the centered resolvent between the readout $r$ and the nonequilibrium defect source $v$.

If $W^\dagger_\perp$ is diagonalizable, with right eigenmodes $\phi_a$ and nonzero eigenvalues $-\mu_a$ satisfying $\Re \mu_a>0$, then
\begin{equation*}
\Delta\chi(\omega)
=
\sum_{a\ge 1}\frac{c_a}{\mu_a-i\omega},
\qquad
c_a=\langle r,\phi_a\rangle_\pi\,\widetilde c_a,
\end{equation*}
for suitable mode amplitudes $\widetilde c_a$ determined by the expansion of $v$. More generally, Jordan blocks produce a finite sum of higher-order poles. In all cases $\Delta\chi$ is a rational meromorphic function of $\omega$ on the complex plane, with no poles on the real axis.

The current representation derived in the End Matter gives
\begin{equation}
v(i)
=
\frac{\beta\zeta}{\pi_i}\sum_j J_{ij}\,[b(i)-b(j)].
\label{eq:SM-current-form}
\end{equation}
Equation \eqref{eq:SM-current-form} shows directly that the defect source is a current--gradient contraction on the network.

\section*{2. Analyticity, symmetry, and vanishing criteria}

Since the state space is finite, $C_{rv}(t)$ is a finite linear combination of exponentially decaying terms multiplied by finite polynomials in $t$. In particular,
\begin{equation*}
C_{rv}\in L^1([0,\infty))\cap L^2([0,\infty)).
\end{equation*}
Therefore $\Delta\chi(\omega)$ is analytic for $\Im \omega>0$ and obeys the standard Kramers--Kronig relations for causal response functions \cite{Kubo1966},
\begin{equation*}
\Re \Delta\chi(\omega)
=
\frac{1}{\pi}\operatorname{p.v.}\!\int_{-\infty}^{\infty}
\frac{\Im \Delta\chi(\omega')}{\omega'-\omega}\,d\omega',
\end{equation*}
\begin{equation*}
\Im \Delta\chi(\omega)
=
-\frac{1}{\pi}\operatorname{p.v.}\!\int_{-\infty}^{\infty}
\frac{\Re \Delta\chi(\omega')}{\omega'-\omega}\,d\omega'.
\end{equation*}

For real observables $r$ and $b$, the kernel $C_{rv}(t)$ is real for real $t$, hence
\begin{equation*}
\Delta\chi(-\omega)=\Delta\chi(\omega)^*,
\end{equation*}
so that
\begin{equation*}
\Re \Delta\chi(\omega)\ \text{is even in }\omega,
\qquad
\Im \Delta\chi(\omega)\ \text{is odd in }\omega.
\end{equation*}

A channel-independent vanishing condition is
\begin{equation*}
W_{\mathrm{asym}}b=0
\quad\Longleftrightarrow\quad
v=0,
\end{equation*}
which implies
\begin{equation*}
\Delta\chi(\omega)\equiv 0
\qquad
\text{for every readout }r.
\end{equation*}
This is stronger than detailed balance. The system may support irreversible steady currents, but the causal mismatch vanishes identically in the perturbation channel generated by $b$ if those currents have no net projection on the $b$-gradients. In current form this means
\begin{equation*}
\sum_j J_{ij}[b(i)-b(j)]=0
\qquad
\text{for all }i.
\end{equation*}

For a fixed readout $r$, one may have $\Delta\chi(\omega)\equiv 0$ even when $v\neq 0$. By \eqref{eq:SM-resolvent}, this happens precisely when $r$ is orthogonal to the entire centered orbit generated by $v$, equivalently
\begin{equation*}
\langle r,(W^\dagger_\perp)^n v\rangle_\pi=0
\qquad
\text{for all }n\ge 0.
\end{equation*}
Thus the mismatch is not a scalar measure of irreversibility alone; it is the part of irreversibility that is both injected by the perturbation coordinate $b$ and visible in the chosen readout $r$.

\section*{3. Low-frequency expansion}

Because $C_{rv}(t)$ decays exponentially, all time moments exist:
\begin{equation*}
M_n\equiv \int_0^\infty dt\,t^n C_{rv}(t)<\infty,
\qquad n=0,1,2,\dots.
\end{equation*}
Therefore $\Delta\chi$ admits a convergent Taylor expansion near $\omega=0$,
\begin{equation}
\Delta\chi(\omega)
=
\sum_{n=0}^\infty \frac{(i\omega)^n}{n!}\,M_n.
\label{eq:SM-lowomega}
\end{equation}
The first terms are
\begin{equation*}
\Delta\chi(\omega)
=
M_0+i\omega M_1-\frac{\omega^2}{2}M_2+O(\omega^3).
\end{equation*}
Hence
\begin{equation*}
\Re \Delta\chi(\omega)=M_0-\frac{\omega^2}{2}M_2+O(\omega^4),
\qquad
\Im \Delta\chi(\omega)=\omega M_1+O(\omega^3).
\end{equation*}
The zero-frequency value
\begin{equation*}
\Delta\chi(0)=M_0
=
\int_0^\infty dt\,C_{rv}(t)
=
\bigl\langle r,(-W^\dagger_\perp)^{-1}v\bigr\rangle_\pi
\end{equation*}
is the static nonequilibrium defect. More generally,
\begin{equation*}
M_n
=
n!\,\bigl\langle r,(-W^\dagger_\perp)^{-(n+1)}v\bigr\rangle_\pi.
\end{equation*}
Equation \eqref{eq:SM-lowomega} shows that the real part of the causal mismatch starts from the static defect, while the imaginary part is linear at small frequency and quantifies the leading low-frequency phase lag of the nonequilibrium correction.

\section*{4. High-frequency asymptotics}

Repeated integration by parts gives the asymptotic expansion
\begin{equation*}
\Delta\chi(\omega)
=
\sum_{n=0}^{m-1}
\frac{\langle r,(W^\dagger)^n v\rangle_\pi}{(-i\omega)^{n+1}}
+
O(|\omega|^{-m-1}),
\qquad |\omega|\to\infty.
\end{equation*}
The first three terms are
\begin{equation*}
\Delta\chi(\omega)
=
\frac{i\langle rv\rangle_\pi}{\omega}
-
\frac{\langle r,W^\dagger v\rangle_\pi}{\omega^2}
-
\frac{i\langle r,(W^\dagger)^2v\rangle_\pi}{\omega^3}
+
O(|\omega|^{-4}).
\end{equation*}
In particular,
\begin{equation*}
|\Delta\chi(\omega)|=O(|\omega|^{-1}).
\end{equation*}
Thus the mismatch necessarily decays at high frequency. The leading $1/\omega$ tail is determined solely by the equal-time overlap $\langle rv\rangle_\pi$. If this overlap vanishes, the tail improves to $O(|\omega|^{-2})$.

Because $\Delta\chi(\omega)$ is not, in general, a positive-real or Herglotz function, neither $\Re\Delta\chi(\omega)$ nor $\Im\Delta\chi(\omega)$ has a definite sign. There is also no general monotonicity in $|\omega|$. Both statements are immediate from the modal representation, since the coefficients $c_a$ may have either sign or complex phase.

\section*{5. Norm identities and generic bounds}

Let
\begin{equation*}
K(t)\equiv \Theta(t)\,C_{rv}(t),
\end{equation*}
so that $\Delta\chi(\omega)$ is the Fourier transform of $K$. Parseval then yields the exact identity
\begin{equation}
\frac{1}{2\pi}\int_{-\infty}^{\infty} d\omega\,|\Delta\chi(\omega)|^2
=
\int_0^\infty dt\,|C_{rv}(t)|^2.
\label{eq:SM-Parseval}
\end{equation}
Equation \eqref{eq:SM-Parseval} makes clear that the integrated mismatch is the squared $L^2$ norm of the causal violation kernel.

Before invoking thermodynamic closures, one already has three generic estimates. The first is the pointwise semigroup bound
\begin{equation}
|\Delta\chi(\omega)|
\le
\int_0^\infty dt\,|C_{rv}(t)|
\le
\frac{\sqrt{\mathrm{Var}(r)\,\mathrm{Var}(v)}}{\lambda}.
\label{eq:SM-generic-pointwise}
\end{equation}
The second is the gap-based integrated bound
\begin{equation}
\frac{1}{2\pi}\int_{-\infty}^{\infty} d\omega\,|\Delta\chi(\omega)|^2
=
\int_0^\infty dt\,|C_{rv}(t)|^2
\le
\frac{\mathrm{Var}(r)\,\mathrm{Var}(v)}{2\lambda}.
\label{eq:SM-generic-L2-gap}
\end{equation}
The third uses the positivity of the bilateral spectral density matrix \cite{Dechant2025FFRIneq},
\begin{equation*}
\mathbb{S}(\omega)=
\begin{pmatrix}
\mathcal{S}_{rr}(\omega) & \mathcal{S}_{rv}(\omega)\\
\mathcal{S}_{vr}(\omega) & \mathcal{S}_{vv}(\omega)
\end{pmatrix}\succeq 0,
\end{equation*}
which implies
\begin{equation*}
|\mathcal{S}_{rv}(\omega)|^2\le \mathcal{S}_{rr}(\omega)\mathcal{S}_{vv}(\omega).
\end{equation*}
Combining this with \eqref{eq:SM-Parseval} and Wiener--Khinchin gives
\begin{equation}
\frac{1}{2\pi}\int_{-\infty}^{\infty} d\omega\,|\Delta\chi(\omega)|^2
\le
\|\mathcal{S}_{rr}\|_\infty\,\mathrm{Var}(v).
\label{eq:SM-generic-L2-sup}
\end{equation}
Substituting the main-text estimate
\begin{equation*}
\mathrm{Var}(v)\le \beta^2\zeta^2\Gamma_b\,\sigma
\le \frac{4\beta^2\zeta^2 D_b}{\pi_{\min}}\,\sigma
\end{equation*}
into \eqref{eq:SM-generic-pointwise}, \eqref{eq:SM-generic-L2-gap}, and \eqref{eq:SM-generic-L2-sup} recovers the fluctuation--dissipation--response inequalities stated in the main text.

\section*{6. Weak-driving scaling}

If all thermodynamic forces are scaled as
\begin{equation*}
F_{ij}=\epsilon f_{ij},
\qquad
\epsilon\ll 1,
\end{equation*}
at fixed symmetric activities $A_{ij}$, then
\begin{equation*}
J_{ij}=A_{ij}\tanh(F_{ij}/2)=O(\epsilon),
\qquad
\sigma=\frac12\sum_{i,j}J_{ij}F_{ij}=O(\epsilon^2),
\end{equation*}
and by \eqref{eq:SM-current-form},
\begin{equation*}
v=O(\epsilon),
\qquad
\Delta\chi(\omega)=O(\epsilon).
\end{equation*}
Consequently,
\begin{equation*}
|\Delta\chi(\omega)|^2=O(\sigma),
\qquad
\frac{1}{2\pi}\int d\omega\,|\Delta\chi(\omega)|^2=O(\sigma).
\end{equation*}
Thus the inequalities are asymptotically sharp in scaling near equilibrium.

The current--traffic step also becomes asymptotically tight. Edgewise,
\begin{equation*}
\frac12 J_{ij}F_{ij}
=
A_{ij}\left(\frac{F_{ij}^2}{4}-\frac{F_{ij}^4}{48}+O(F_{ij}^6)\right),
\end{equation*}
whereas
\begin{equation*}
\frac{J_{ij}^2}{A_{ij}}
=
A_{ij}\left(\frac{F_{ij}^2}{4}-\frac{F_{ij}^4}{24}+O(F_{ij}^6)\right).
\end{equation*}
Summing over edges gives
\begin{equation*}
\sigma-\sum_{i,j}\frac{J_{ij}^2}{A_{ij}}
=
O(\epsilon^4).
\end{equation*}
Hence the relative slack of the thermodynamic closure is $O(\epsilon^2)$.

\section*{7. Ring case: exact mode analysis and near saturation}

Consider an $N$-state ring, with indices understood modulo $N$, and uniform clockwise and counterclockwise rates
\begin{equation}
W_{i+1,i}=k\,e^{F/2},
\qquad
W_{i-1,i}=k\,e^{-F/2}.
\label{eq:SM-ring-rates}
\end{equation}
The stationary distribution is uniform,
\begin{equation*}
\pi_i=\frac{1}{N}.
\end{equation*}
For a wavenumber
\begin{equation*}
q=\frac{2\pi m}{N},
\qquad
m=1,\dots,\left\lfloor\frac{N-1}{2}\right\rfloor,
\end{equation*}
introduce the normalized real Fourier pair
\begin{equation*}
c_q(i)=\sqrt{2}\cos(qi),
\qquad
s_q(i)=\sqrt{2}\sin(qi),
\end{equation*}
which satisfy
\begin{equation*}
\langle c_q^2\rangle_\pi=\langle s_q^2\rangle_\pi=1,
\qquad
\langle c_q,s_q\rangle_\pi=0.
\end{equation*}
A direct calculation from \eqref{eq:SM-ring-rates} gives the invariant two-dimensional mode subspace
\begin{equation*}
W_{\mathrm{sym}}c_q=-\mu_q c_q,
\qquad
W_{\mathrm{sym}}s_q=-\mu_q s_q,
\end{equation*}
\begin{equation*}
W_{\mathrm{asym}}c_q=-\Omega_q s_q,
\qquad
W_{\mathrm{asym}}s_q=\Omega_q c_q,
\end{equation*}
with
\begin{equation}
\mu_q=2k\cosh(F/2)\,[1-\cos q],
\qquad
\Omega_q=2k\sinh(F/2)\,\sin q.
\label{eq:SM-muOmega}
\end{equation}
The symmetric part sets the decay rate, while the antisymmetric part rotates the mode pair.

Choose the perturbation observable
\begin{equation*}
b=c_q.
\end{equation*}
Then
\begin{equation*}
v=2\zeta\beta W_{\mathrm{asym}}b
=
-2\zeta\beta\,\Omega_q\,s_q.
\end{equation*}
If we also choose the readout
\begin{equation*}
r=s_q,
\end{equation*}
then the readout is exactly aligned with the defect source, and
\begin{equation*}
\mathrm{Var}(r)=1,
\qquad
\mathrm{Var}(v)=4\zeta^2\beta^2\Omega_q^2.
\end{equation*}
The violation kernel is then
\begin{equation}
C_{rv}(t)
=
-2\zeta\beta\,\Omega_q\,e^{-\mu_q t}\cos(\Omega_q t),
\qquad t\ge 0.
\label{eq:SM-ring-C}
\end{equation}
Its one-sided Fourier transform can be written in either of the equivalent forms
\begin{equation*}
\Delta\chi_q(\omega)
=
-2\zeta\beta\,\Omega_q\,
\frac{\mu_q-i\omega}{(\mu_q-i\omega)^2+\Omega_q^2},
\end{equation*}
or
\begin{equation}
\Delta\chi_q(\omega)
=
-\zeta\beta\,\Omega_q
\left[
\frac{1}{\mu_q-i(\omega-\Omega_q)}
+
\frac{1}{\mu_q-i(\omega+\Omega_q)}
\right].
\label{eq:SM-ring-chi-poles}
\end{equation}
Equations \eqref{eq:SM-ring-C}--\eqref{eq:SM-ring-chi-poles} show explicitly how the mismatch is produced by a competition between decay at rate $\mu_q$ and reversible phase rotation at rate $\Omega_q$.

At zero frequency,
\begin{equation*}
\Delta\chi_q(0)
=
-2\zeta\beta\,\Omega_q\,
\frac{\mu_q}{\mu_q^2+\Omega_q^2}.
\end{equation*}
If $q=2\pi/N$ is the slowest nonzero mode, then $\lambda=\mu_q$ and the generic pointwise bound \eqref{eq:SM-generic-pointwise} becomes
\begin{equation*}
|\Delta\chi_q(\omega)|
\le
\frac{\sqrt{\mathrm{Var}(r)\,\mathrm{Var}(v)}}{\lambda}
=
\frac{2\zeta\beta\,|\Omega_q|}{\mu_q}.
\end{equation*}
At $\omega=0$, the exact saturation ratio is therefore
\begin{equation}
R_{\infty}
\equiv
\frac{|\Delta\chi_q(0)|}{\sqrt{\mathrm{Var}(r)\,\mathrm{Var}(v)}/\lambda}
=
\frac{\mu_q^2}{\mu_q^2+\Omega_q^2}
=
\frac{1}{1+(\Omega_q/\mu_q)^2}.
\label{eq:SM-pointwise-ratio}
\end{equation}
Using \eqref{eq:SM-muOmega},
\begin{equation*}
\frac{\Omega_q}{\mu_q}
=
\tanh(F/2)\,\cot(q/2).
\end{equation*}
Thus the pointwise bound is near saturation when
\begin{equation*}
\tanh(F/2)\,\cot(q/2)\ll 1.
\end{equation*}
For the slowest mode $q=2\pi/N$, this requires
\begin{equation*}
F\ll 2\tan(\pi/N)
\end{equation*}
in the weak-driving regime. Mode purity alone is therefore not sufficient: if $N$ is large at fixed nonzero drive, then $\Omega_q/\mu_q\sim F/q$ grows and the oscillatory phase in \eqref{eq:SM-ring-C} strongly suppresses the one-sided transform.

The integrated norm can also be evaluated exactly from \eqref{eq:SM-ring-C}:
\begin{equation*}
\frac{1}{2\pi}\int_{-\infty}^{\infty} d\omega\,|\Delta\chi_q(\omega)|^2
=
\int_0^\infty dt\,|C_{rv}(t)|^2
=
\zeta^2\beta^2\Omega_q^2
\left(
\frac{1}{\mu_q}
+
\frac{\mu_q}{\mu_q^2+\Omega_q^2}
\right).
\end{equation*}
If again $q=2\pi/N$ so that $\lambda=\mu_q$, then the gap-based integrated bound \eqref{eq:SM-generic-L2-gap} reads
\begin{equation*}
\frac{1}{2\pi}\int d\omega\,|\Delta\chi_q(\omega)|^2
\le
\frac{\mathrm{Var}(r)\,\mathrm{Var}(v)}{2\lambda}
=
\frac{2\zeta^2\beta^2\Omega_q^2}{\mu_q},
\end{equation*}
and the exact integrated saturation ratio is
\begin{equation}
R_2
\equiv
\frac{\frac{1}{2\pi}\int d\omega\,|\Delta\chi_q(\omega)|^2}
{\mathrm{Var}(r)\,\mathrm{Var}(v)/(2\lambda)}
=
\frac12
\left(
1+\frac{\mu_q^2}{\mu_q^2+\Omega_q^2}
\right).
\label{eq:SM-integrated-ratio}
\end{equation}
Hence
\begin{equation*}
R_2=1-\frac12\left(\frac{\Omega_q}{\mu_q}\right)^2+O\!\left(\frac{\Omega_q^4}{\mu_q^4}\right)
\qquad
\text{as }\frac{\Omega_q}{\mu_q}\to 0,
\end{equation*}
so the integrated gap-based inequality is also analytically near saturation in the weak-rotation regime.

The ring also illustrates the thermodynamic closure. The stationary current and symmetric activity on each oriented edge are
\begin{equation*}
J=\frac{2k}{N}\sinh(F/2),
\qquad
A=\frac{2k}{N}\cosh(F/2).
\end{equation*}
Since the force on each clockwise edge is $F$ and on each counterclockwise edge is $-F$, one finds
\begin{equation*}
\sigma=NJ\,F=2kF\sinh(F/2),
\end{equation*}
and
\begin{equation*}
\sum_{i,j}\frac{J_{ij}^2}{A_{ij}}
=
2N\,\frac{J^2}{A}.
\end{equation*}
Therefore
\begin{equation}
\frac{\sum_{i,j}J_{ij}^2/A_{ij}}{\sigma}
=
\frac{2\tanh(F/2)}{F}
=
1-\frac{F^2}{12}+O(F^4).
\label{eq:SM-ring-current-traffic-ratio}
\end{equation}
Equation \eqref{eq:SM-ring-current-traffic-ratio} shows that the thermodynamic current--traffic step is asymptotically tight as $F\to 0$.

Taken together, \eqref{eq:SM-pointwise-ratio}, \eqref{eq:SM-integrated-ratio}, and \eqref{eq:SM-ring-current-traffic-ratio} provide an explicit analytic near-saturation regime for the causal fluctuation--dissipation--response inequalities on a solvable Markov network. The mechanism is transparent. One needs mode purity, so that $r$ and $v$ live on the same two-dimensional invariant subspace; slow reversible decay, so that $\lambda=\mu_q$ is the relevant timescale; weak phase rotation, so that $\Omega_q/\mu_q\ll 1$ and the one-sided transform loses little coherence; and weak driving, so that the thermodynamic current--traffic closure is close to equality. Exact nontrivial saturation remains excluded away from the zero-violation limit, because any finite phase rotation lowers the one-sided transform below the purely relaxational envelope, and any finite force leaves a strictly subleading slack in the current--traffic comparison.

\end{document}